\documentclass[final,5p,times]{elsarticle}
\usepackage{graphicx}

\usepackage{amssymb}

\usepackage{xspace}
\usepackage{textcomp}
\usepackage{booktabs}
\usepackage{graphicx}
\usepackage{caption}
\usepackage{subcaption}
\usepackage{hyperref}
\usepackage{siunitx}

\newcommand{\tp}{\textsc{TelePix}\xspace}
\newcommand{\pmos}{{PMOS}\xspace}
\newcommand{\nmos}{{NMOS}\xspace}
\newcommand{\cmos}{{CMOS}\xspace}
\newcommand{\rzeron}{{\textsc{Run2020}-NMOS}\xspace}
\newcommand{\rzerop}{{\textsc{Run2020}-PMOS}\xspace}
\newcommand{\ronec}{{\textsc{Run2021}-CMOS}\xspace}
\newcommand{\ronen}{{\textsc{Run2021}-NMOS}\xspace}

\newcommand{\hb}{\textsc{HitBus}\xspace}

\makeatletter
\def\ps@pprintTitle{%
	\let\@oddhead\@empty
	\let\@evenhead\@empty
	\def\@oddfoot{\textit{Preprint submitted to Proceedings of the 15th Pisa Meeting on Advanced Detectors}\hfill}%
	\let\@evenfoot\@oddfoot}

\makeatother
\journal{Nuclear Instruments and Methods A}

\begin{document}
	\begin{frontmatter}
		\title{\tp\,- A fast region of interest trigger and timing layer for the {EUDET} Telescopes}
		\author[a]{Heiko~Augustin}		
		\author[a]{Sebastian~Dittmeier}
		\author[b]{Jan~Hammerich}
		\author[c]{Adrian~Herkert}
		\author[c]{Lennart~Huth\corref{cor1}}
		\ead{lennart.huth@desy.de}
		\author[a]{David~Immig}
		\author[d]{Ivan Peri\'{c}}
		\author[a]{Andr\'{e}~Sch\"oning}
		\author[c]{Adriana~Simancas}
		\author[c]{Marcel~Stanitzki}
		\author[a]{Benjamin~Weinl\"ader}
		\cortext[cor1]{corresponding author}

		\address[a]{Physikalisches Institut der Universit\"at Heidelberg, INF 226, 69120 Heidelberg, Germany}
		\address[b]{University of Liverpool, Liverpool L69 3BX, United Kingdom}
		\address[c]{Deutsches Elektronen-Synchrotron DESY, Notkestr. 85, 22607 Hamburg, Germany}
	    \address[d]{Institut f\"ur Prozessdatenverarbeitung und Elektronik, KIT, Hermann-von-Helmholtz-Platz 1, 76344 Eggenstein-Leopoldshafen, Germany}
	

		\begin{abstract}
			Test beam facilities are essential to study the response of novel detectors to particles. 
			At the DESY II Test Beam facility, users can test their detectors with an electron beam with a momentum from \SI[parse-numbers=false]{1-6}{GeV}. To track the beam particles, EUDET-style telescopes are provided in each beam area. 
			They provide excellent spatial resolution, but the time resolution is limited by the rolling shutter architecture to a precision of approximately \SI{230}{\micro s}. Since the demand on particle rates - and hence track multiplicities - is increasing  timing is becoming more relevant.
			 DESY foresees several upgrades of the telescopes. 
			\tp is an upgrade project to provide track timestamping with a precision of   better than \SI{5}{ns} and a configurable region of interest to trigger the telescope readout. 
			Small scale prototypes have been characterised in laboratory and test beam measurements.  
			Laboratory tests with an injection corresponding to 2300 electrons show a S/N of above 20.
			Test beam characterization shows  efficiencies of above \SI{99}{\percent} over a threshold range of more than 100mV and time resolutions of \SI{2.4}{ns} at low noise rates. 
		\end{abstract}
	\end{frontmatter}

	\section{Introduction}
	Test beam measurements provide conditions very close to applications in experiments and are hence crucial for sensor R\&D.  
	DESY II \cite{DIENER2019265} provides three independent user beam lines with electrons/positrons  ranging from \SI[parse-numbers=false]{1-6}{GeV}.
	Each line is equipped with a EUDET-style \cite{jansen2016} reference tracking telescope based on MIMOSA-26 sensors \cite{baudot2009,huguo2010}.
	They provide excellent position resolution with an integration time of \SI{230}{\micro s}. 
	At typical operation conditions, this creates events with up to 6 particles passing the telescope, causing ambiguities. Additionally, small devices under test (DUT) are likely to not be penetrated by the particle triggered on, which causes inefficient data taking.
	
	A new telescope plane, \tp, is foreseen to solve both issues by providing a precise time stamp and a fast region of interest trigger output on arbitrary pixel arrangements. A similar hybrid approach with less precise timing is presented in \cite{FE-I4}.

	\section{\tp Prototypes}
	The prototypes are designed in  a \SI{180}{nm} HV-CMOS process and profit from more than a decade of research and experience in this process \cite{PERIC2007876,SCHIMASSEK2021164812}. 
	They feature a matrix of \SI[parse-numbers=false]{29\times124}{pixels} at a pitch of \SI[parse-numbers=false]{165x25}{\micro\meter\squared} and are operated in a data driven mode with a column drain readout logic.
    A configurable fast digital hit-OR output (\hb) of all unmasked pixels is foreseen as a region of interest trigger of the reference telescopes.  This provides the flexibility to trigger on any subset of pixels, to optimally match the DUT.
    Readout is handled by the Mu3e pixel DAQ \cite{TWEPP2017,Dittmeier2018}.
	Four different detector flavours from two submissions are compared.
	A chip split into two parts with \nmos \& \pmos  transistors at the amplifier input was submitted in 2020. They are referred to as \rzeron and \rzerop in the following.
	Improved biasing alongside \nmos and \cmos inputs are realized in two chips in  a later submission (Run2021), referred to as \ronen and \ronec.
	For the presented studies, the chips are operated at a core frequency of \SI{125}{MHz} and  time stamp precision of \SI{4}{ns}. No threshold trimming or pixel masking is applied. All samples have been thinned to \SI{100}{\micro m}.

	\subsection*{Laboratory Characterization}
\begin{figure}[h]
	\centering
\includegraphics[width=0.8\columnwidth]{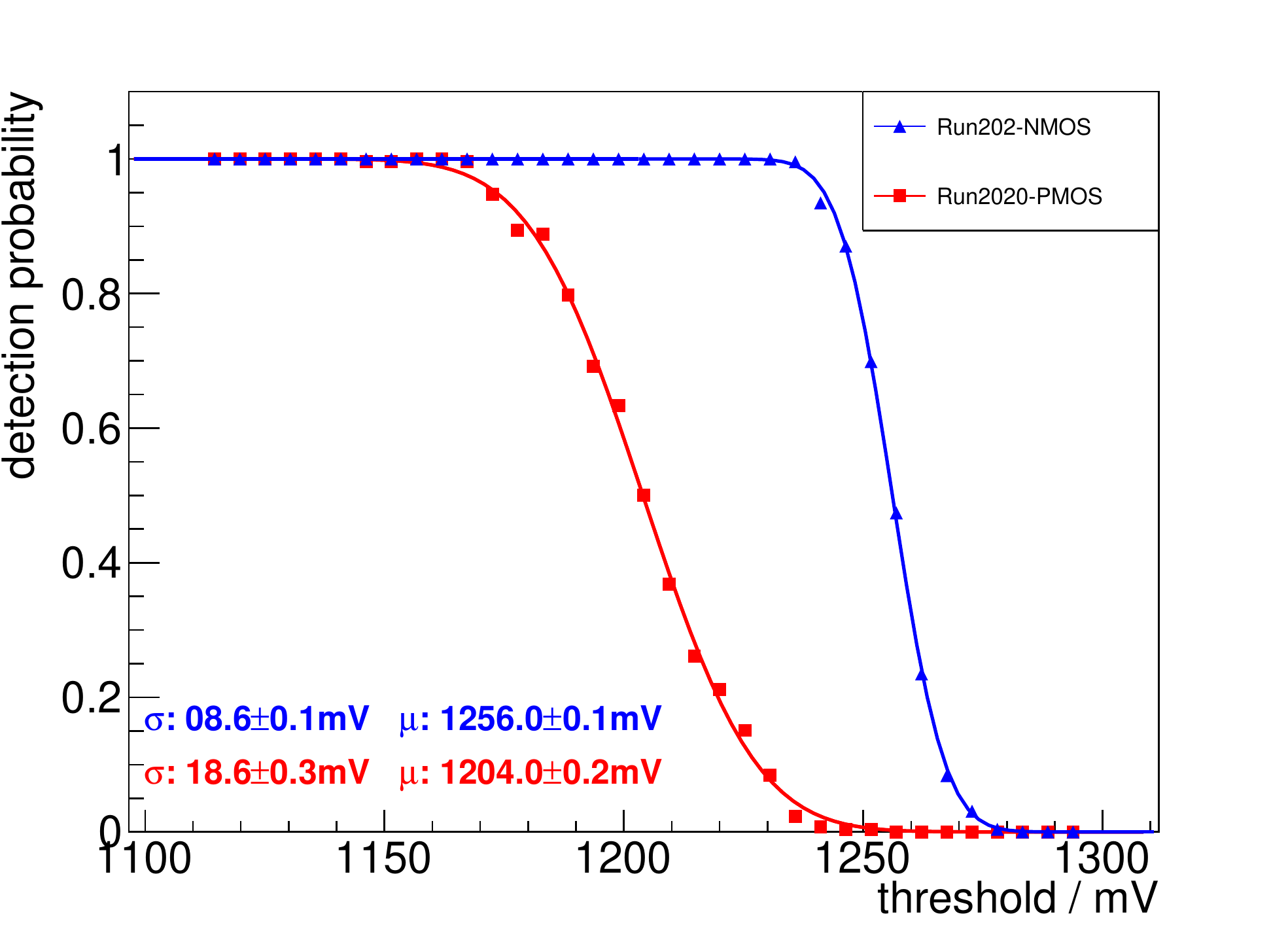}
\caption{Exemplary response to the charge injection for  \rzeron (blue triangles) and \rzerop (red squares). An s-curve is fitted to the data. The baseline is at \SI{1}{V}. \label{s-curve}}
\end{figure}

\begin{figure}[bh]
		\centering
	\begin{subfigure}{0.49\columnwidth}
		\includegraphics[width=\textwidth]{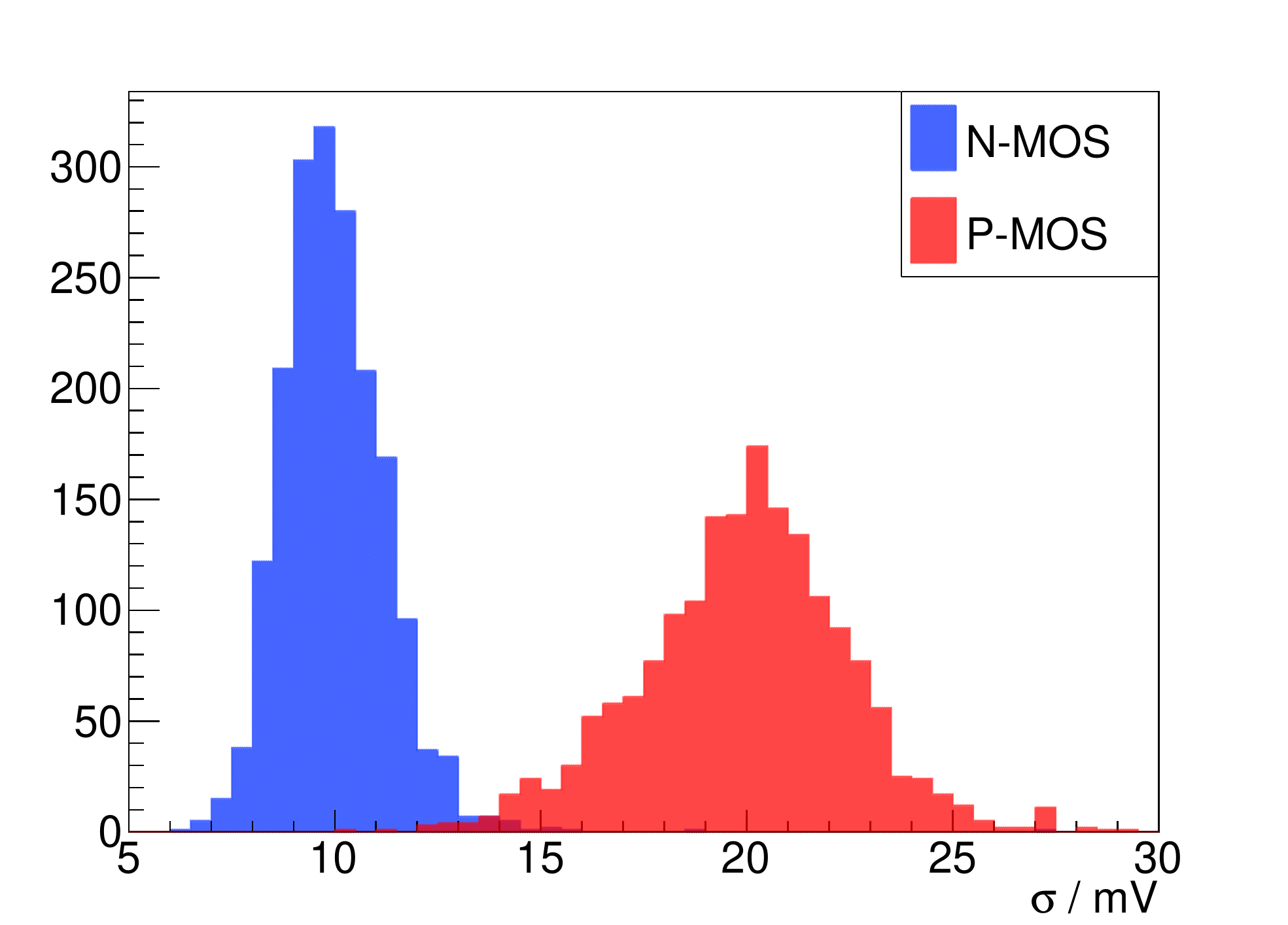}
		\caption{sigma of the s-curve}		
	\end{subfigure}
	\hfill
	\begin{subfigure}{0.49\columnwidth}
		\includegraphics[width=\textwidth]{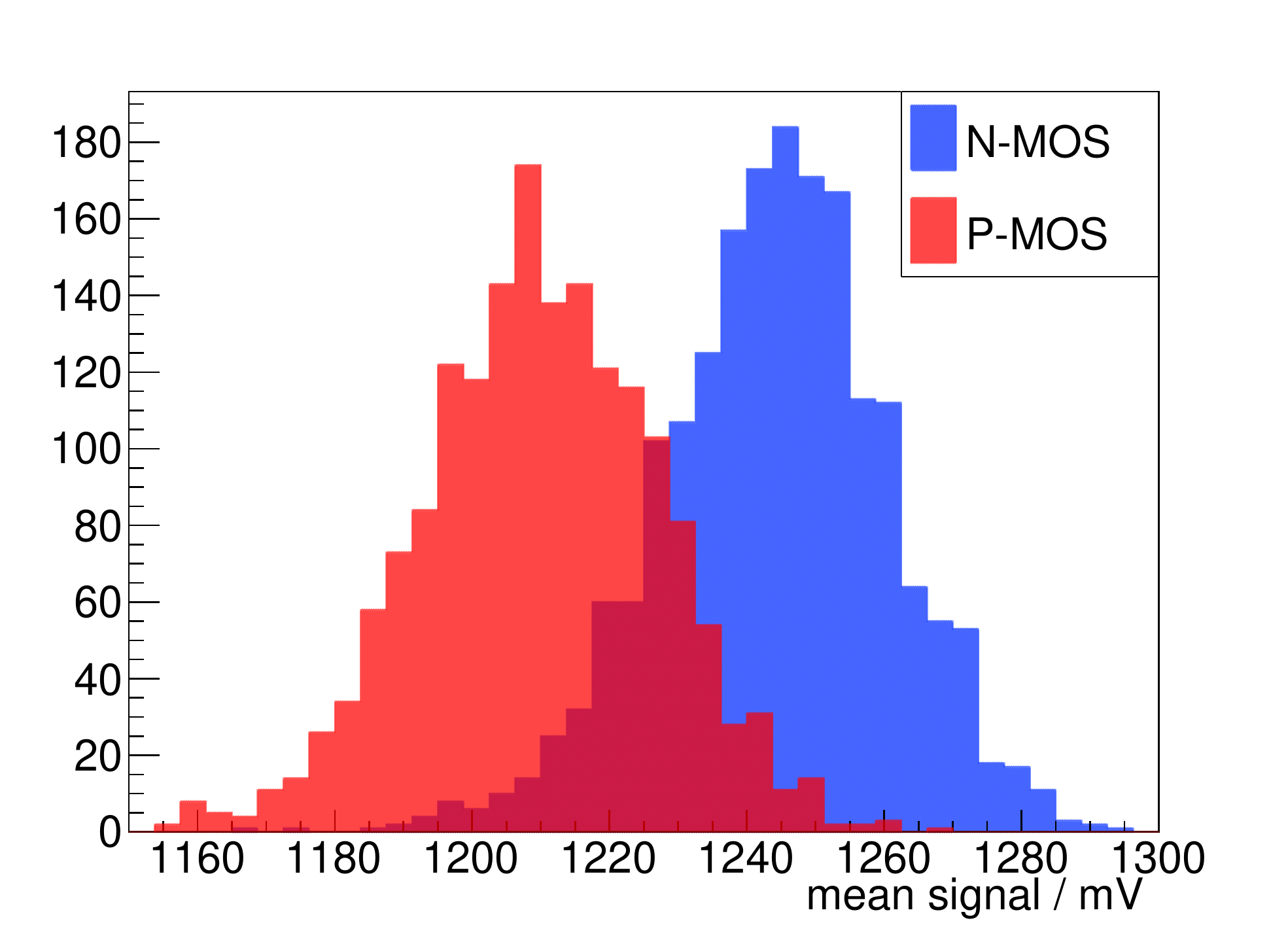}
		\caption{Mean signal}
	\end{subfigure}
	\caption{Results of the injection scans. \rzerop(\rzeron) in red (blue).\label{sigma}}
\end{figure}

	The first submission was studied in the laboratory using the charge injection feature of the chip (400mV, $\approx$2300 electrons) and analysing the response of the fast \hb.
	Figure~\ref{s-curve} shows the fraction of detected injections for the \rzeron and \rzerop type, exemplary for a single pixel.
	The \nmos generates a larger signal with less noise at the used settings. 
	In figure~\ref{sigma}, the sigma and mean of the s-curve fits are histogrammed for all pixels  of the two types.  A signal to noise ratio of above 20 was determined for the \nmos.
	
\subsection*{Test Beam Characterization}
	The \tp prototypes have been characterized in test beam campaigns at the DESY II facility. 
	The test setup, corryvreckan framework \cite{Dannheim_2021} based analysis as well as first measurements are presented in \cite{huth_vci}. 
	
\begin{figure}[h]
		\centering
		\begin{subfigure}{0.8\columnwidth}
		\includegraphics[width=\textwidth]{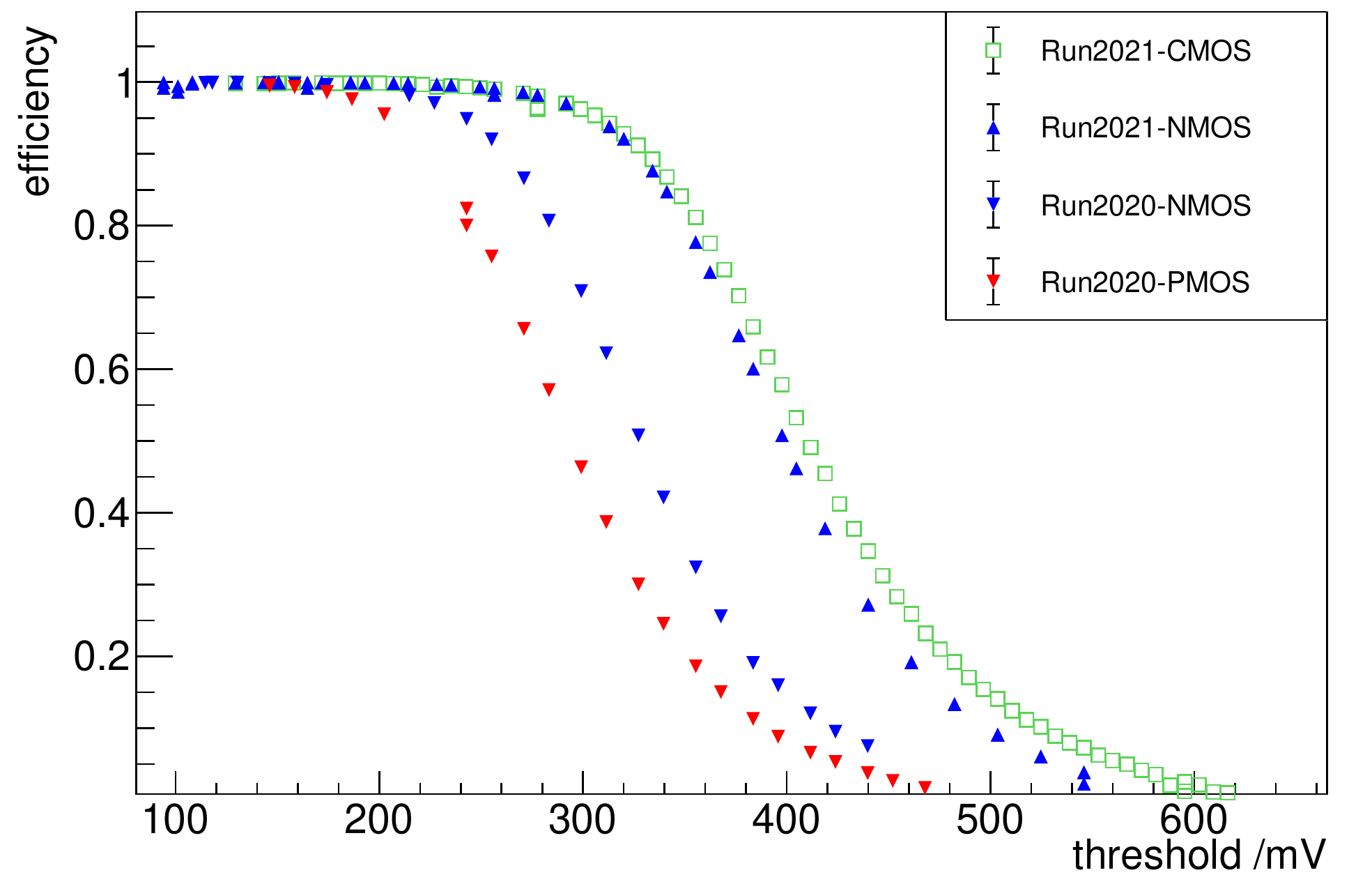}
\caption{Efficiency as a function of the detection threshold. Run2020 results from\cite{huth_vci}. Statistical errors are too small to be seen. \label{efficiency}}		\end{subfigure}
\hfill
\begin{subfigure}{0.6\columnwidth}
\includegraphics[width=\textwidth]{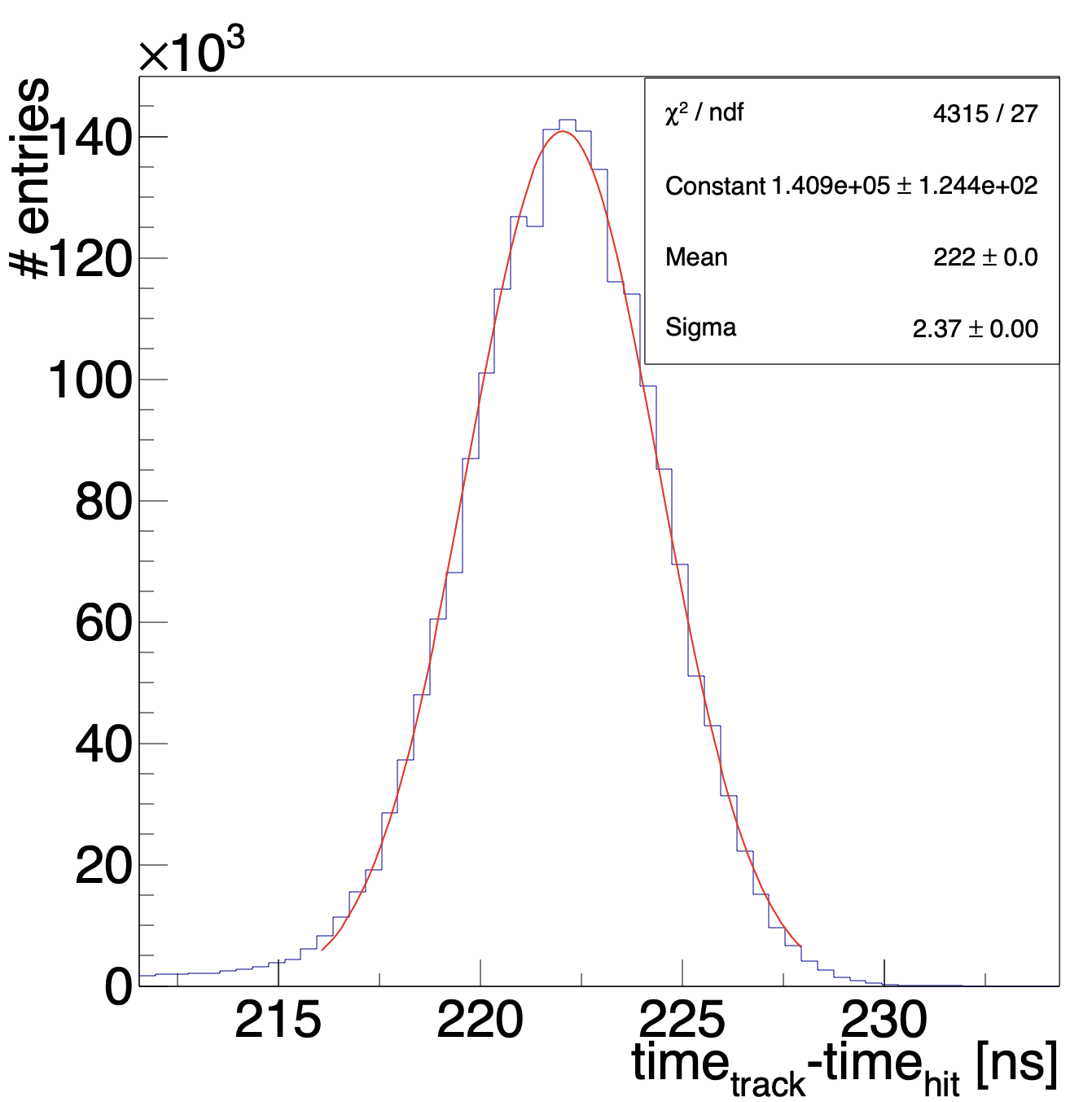}
\caption{Time resolution of the \cmos chip. Gaussian fit to the core of the distribution.\label{time_res}}
\end{subfigure}
\caption{Efficiency vs threshold of all tested prototypes and time resolution of the \ronec. The samples are biased with \SI{-70}{V}.}
	\end{figure}

	Figure~\ref{efficiency} summarizes the efficiency as a function of the threshold for all tested samples.  It can be  seen, that the efficiency plateau of the \rzerop is smaller than for the \rzeron. 
	With the improvements in Run2021, a significant increase in the efficiency plateau is observed. 
	\ronec has a slightly wider high efficiency region. The noise levels of all prototypes are similar.  

	The time resolution of the \ronec at a  low detection threshold of \SI{108}{mV}, with respect to a fast LySo crystal connected to a SiPM has been measured in figure~\ref{time_res}. For HV-CMOS an unprecedented resolution of below \SI{2.4}{ns} is determined, which is significantly better than the requirements. 
	The {\it region of interest trigger feature} has been evaluated in \cite{huth_vci}. It had an absolute delay of $\approx$\SI{22}{ns} and a jitter of  $\approx$\SI{3.8}{ns} for \rzeron, including  pixel-to-pixel fluctuations within a column.

	\section{Summary \& Outlook}
	The two \tp prototype submissions have been successfully tested in beam and laboratory measurements. The Run2021 submission further  improves the already excellent performance of the sensor. Threshold regions of above 100~mV with $>$\SI{99}{\percent} hit detection efficiency have been determined alongside excellent time resolutions of \SI{2.4}{ns}. \tp is meeting all requirements to serve as trigger and timing plane at the DESY II test beam. The region of interest trigger feature has been tested and is fully functional and already serving first users.
	
	A chip with an active area of \SI[parse-numbers=false]{2\times1}{cm\squared}  in the \ronec flavour has been submitted and is expected to be delivered in December 2022. Due to the experience in large scale sensor design in the process, we do not expect major issues in the submission. Deployment of the full scale sensor to all telescopes is expected during the first half of 2023. 
	With the \tp upgrade timing/triggering performance of the  EUDET-style telescopes will improve significantly and allow for efficient future test beams.
		\section*{Acknowledgements}
						The measurements leading to these results have been performed at the Test Beam Facility at DESY Hamburg (Germany), a member of the Helmholtz Association (HGF). This project has received funding from the European Union’s Horizon 2020 Research and Innovation programme under  GA no 101004761.  We acknowledge support by the Deutsche Forschungsgemeinschaft (DFG, German Research Foundation) under Germany’s Excellence Strategy – EXC 2121 "Quantum Universe“ – 390833306. 
	\bibliography{bib}
\end{document}